\documentclass[preprint2]{aastex}

\usepackage{graphicx}
\DeclareGraphicsExtensions{.ps}
\shorttitle{\indent \def Solar Tornadoes Observed with IRIS} \shortauthors{Yang et al.}

\begin{document}

\title{Two Solar Tornadoes Observed with the Interface Region Imaging Spectrograph}

\author{Zihao Yang\altaffilmark{1}, Hui Tian\altaffilmark{1}, Hardi Peter\altaffilmark{2}, Yang Su\altaffilmark{3}, Tanmoy Samanta\altaffilmark{1}, Jingwen Zhang\altaffilmark{1}, Yajie Chen\altaffilmark{1}}
\altaffiltext{1}{School of Earth and Space Sciences, Peking University, 100871 Beijing, China; huitian@pku.edu.cn}
\altaffiltext{2}{Max-Planck Institute for Solar System Research, Justus-von-Liebig-Weg 3, D-37077 G{\"o}ttingen, Germany}
\altaffiltext{3}{Key Laboratory of Dark Matter and Space Astronomy, Purple Mountain Observatory, Chinese Academy of Sciences, Nanjing, Jiangsu, China}

\begin{abstract}
The barbs or legs of some prominences show an apparent motion of rotation, which are often termed solar tornadoes. It is under debate whether the apparent motion is a real rotating motion, or caused by oscillations or counter-streaming flows. We present analysis results from spectroscopic observations of two tornadoes by the Interface Region Imaging Spectrograph. Each tornado was observed for more than 2.5 hours. Doppler velocities are derived through a single Gaussian fit to the Mg~{\sc{ii}}~k~2796\AA{}~and Si~{\sc{iv}}~1393\AA{}~line profiles. We find coherent and stable redshifts and blueshifts adjacent to each other across the tornado axes, which appears to favor the interpretation of these tornadoes as rotating cool plasmas with temperatures of $10^4$ K-$10^5$ K. This interpretation is further supported by simultaneous observations of the Atmospheric Imaging Assembly on board the Solar Dynamics Observatory, which reveal periodic motions of dark structures in the tornadoes. Our results demonstrate that spectroscopic observations can provide key information to disentangle different physical processes in solar prominences.
\end{abstract}


\section{Introduction}
Solar prominences are cool and dense materials embedded in the hot corona above the solar limb. On the disk, they often show up as dark filament structures in H$\alpha$ and Extreme ultraviolet observations. Morphologically, solar prominences consist of spines, legs, and barbs \citep[e.g.,][]{Martin1998,Parenti2014,Shen2015}. It has long been known that some parts of a subset of prominences show a sign of rotation. Edison Pettit first noticed this type of prominences and categorized them as tornadoes \citep{Pettit1943}. More recently, the term tornado has also been used to describe some other rotating solar phenomena including the macrospicules with rotating motions \citep{Pike&Mason1998,Curdt&Tian2011}, chromospheric swirls which may channel energy to the corona \citep{Wedemeyer&Voort2009,Wedemeyer2012,Yang2015} and fast-evolving tornado-like erupting structures \citep{Chen2017}. Other rotating structures such as coronal cyclones rooted in the rotating network magnetic field \citep{Zhang&Liu2011} and rotating jet-like structures \citep[e.g.,][]{Shen2011,Shen2012} have also received attention in recent years. In this paper, the term solar tornadoes refer to the barbs or legs of some prominences that show an apparent rotating motion.

Many solar tornadoes have been identified since the launch of the Solar Dynamics Observatory \citep[SDO,][]{Pesnell2012} and Hinode. Through observations of the Atmospheric Imaging Assembly \citep[AIA,][]{Lemen2012} on board SDO, \cite{Li2012} found the formation of a solar tornado and its relationship with the coronal cavity above. \cite{Su2012} found that the rotating solar tornadoes observed with SDO/AIA have a connection with prominences. The possible relationship between a solar tornado and CMEs/flares from the surrounding active region has also been reported \citep{Panesar2013}. \cite{Wedemeyer2013} performed a statistical analysis of tornadoes using SDO/AIA observations and found that these solar tornadoes, which they called giant tornadoes, have a close relationship with prominences, share some similar characteristics with barbs and may serve as the plasma source of some prominences. Attempts of plasma diagnostics and magnetic field measurements have also been made for solar tornadoes  \citep[e.g.,][]{Levens2016b,Gonzalez2016,Levens2017}. A theoretical investigation by \cite{Luna2015} suggests that the cool and dense plasma in these tornadoes can be supported by the Lorentz force if the associated magnetic structure is sufficiently twisted or strong poloidal flows are present.

There are different interpretations for the apparent rotating motion of solar tornadoes, including rotation, oscillation and counter-streaming. \cite{Su2014} and \cite{Levens2015} analyzed a solar tornado observed by the EUV Imaging Spectrometer \citep[EIS,][]{Culhane2007} on board Hinode and found split red and blue shifts across the tornado. They interpreted this result as the rotational motion of plasma in the tornado. However, the emission lines they used are formed at coronal temperatures. Thus, this observational signature can only be explained as the rotation of the hot coronal plasma surrounding the tornado and might not reflect the motion of the cool tornado materials. On the contrary, some researchers believe that the apparent rotating motion is not real rotation. For instance, \cite{Panasenco2014} mentioned that such apparent motion is a visual illusion caused by the projection of oscillating plasma or counter-streaming flows. Using observations of two tornadoes with the Interface Region Imaging Spectrograph \citep[IRIS,][]{DePontieu2014}, \cite{Levens2016a} found no evidence of rotation through an analysis of the Mg~{\sc{ii}}~k~Doppler shift. More recently, \cite{Schmieder2017} analyzed a solar tornado observed with H$\alpha$ and found an alternation between redshifted and blueshifted regions, revealing a possible irregular periodicity of about 1 hour. Such behavior is not supposed to be caused by rotation. It is unclear how frequent such oscillations are in solar tornadoes.

Obviously, more high-resolution observations, especially spectroscopic observations, are required to disentangle these different scenarios for solar tornadoes. Here we present analysis results from IRIS observations of two solar tornadoes. IRIS has several strong emission lines formed in the temperature range of $10^4$ K-$10^5$ K. Observations with these lines provide information about the motion of the cool tornado materials instead of the hot coronal plasmas in the surrounding environment. Both of our IRIS observations lasted for more than 2.5 hours, which is sufficient to detect possible oscillations with a period of $\sim$1 hour \citep{Schmieder2017}. Our analysis of the IRIS spectra appears to support the scenario of rotation for these two tornadoes.

\section{Observations}

Since its launch in 2013, IRIS has been obtaining spectra and slit-jaw images (SJIs) with a high spatial resolution and cadence. The slit-jaw imager has four passbands sampling mainly the C~{\sc{ii}}~1334\AA{}/1335\AA{}~lines, the Si~{\sc{iv}}~1393\AA{}/1400\AA{}~lines, the Mg~{\sc{ii}}~k~2796\AA{}~line and the photospheric continuum around 2832\AA, respectively. The spectrograph has two wavelength bands in the far ultraviolet (FUV, 1332\AA{}-1358\AA{} and 1389\AA{}-1407\AA{}) and one wavelength band in the near ultraviolet (NUV, 2783\AA{}-2834\AA{}) \citep{DePontieu2014}. Here we analyze two tornado-like prominences (T1 and T2) observed by IRIS.

\subsection{First tornado (T1)}

The first solar tornado was observed with IRIS using a sit-and-stare observation mode from 5:50 to 8:21 UT on 2014 April 9. The pointing coordinate of the telescope was (-638$^{\prime\prime}$, -764$^{\prime\prime}$) and the slit was rolled by -48 degrees. The spatial pixel size was 0.167$^{\prime\prime}$. With an exposure time of 8 s, 960 exposures were taken at a cadence of 9.4 s. Slit-jaw images in the filters of 1400\AA{} and 2796\AA{} were taken alternatively, with a field of view (FOV) of $119^{\prime\prime}\times119^{\prime\prime}$ and a cadence of 19 s for each passband. We use calibrated level 2 data for our analysis.  Dark current subtraction, flat field, and geometrical corrections have all been applied to the data \citep{DePontieu2014}.

We mainly use the Si~{\sc{iv}}~1393.755\AA{}~and Mg~{\sc{ii}}~k~2796.347\AA{}~lines for spectral analysis, since the formation temperatures of these two strong lines fall in the temperature regime of prominences ($\sim$ 0.1 MK for Si~{\sc{iv}}~and $\sim$ 0.01 MK for Mg~{\sc{ii}}). These two lines are among the strongest spectral lines in IRIS spectrum, and thus they provide an excellent tool for prominence diagnostics. We take the average line centers above the solar limb as the rest wavelengths and calculate the Doppler shift of each line at each spatial pixel of the tornado.

\subsection{Second tornado (T2)}

The second observation was a 16-step raster scan made from 00:09 to 03:40 UT on 2017 March 12. The pointing coordinate of the telescope was (702$^{\prime\prime}$, 748$^{\prime\prime}$). The spatial pixel size was 0.167$^{\prime\prime}$. The FOV of the spectral observation was $30^{\prime\prime}\times174^{\prime\prime}$ and each raster covered sixteen 2$^{\prime\prime}$ steps. The step cadence was 31.6 s, and the raster cadence was 502 s. Slit-jaw images in the filters of 1400\AA{} and 2796\AA{} were taken alternatively. The FOV of SJIs was $167^{\prime\prime}\times174^{\prime\prime}$ and the cadence was 63 s for each passband. Also, the level 2 data are used.

The same Si~{\sc{iv}}~and Mg~{\sc{ii}}~k~lines are used for our spectral analysis. The orbital variation (both the thermal component and the spacecraft velocity component) has been subtracted in the level 2 data \citep{Tian2014}. However, we find that there is still a residual orbital variation. To avoid misinterpretation of the data, we derive the average line centroid of the Ni~{\sc{i}}~2799.474\AA{}~line on the solar disk as a function of time through a Gaussian fit. The average value is then subtracted from this trend. Afterwards, the resultant relative variation is subtracted from the Doppler shift of the Mg~{\sc{ii}}~k~2796.347\AA{}~line. Since the residual (in the unit of \AA{}) is anti-correlated between NUV and FUV lines \citep{DePontieu2014}, we can also apply the result derived from the Ni~{\sc{i}}~2799.474\AA{}~line to the Si~{\sc{iv}}~1393.755\AA{}~line. For absolute wavelength calibration in the FUV wavelength band, we assume that the Fe~{\sc{ii}}~1392.817\AA{}~line has no Doppler shift on average on the solar disk. For absolute wavelength calibration in the NUV band, several strong absorption lines are assumed to be at rest on the solar disk.

We also use the SDO/AIA data to study the second tornado. We have analyzed images taken in the AIA 193\AA{} passband. The pixel size of AIA 193\AA{} images is $\sim0.6^{\prime\prime}$, and the cadence is 12 s.

\section{Results and Discussion}
\subsection{First tornado (T1)}
\begin{figure*}
\centering {\includegraphics[width=\textwidth]{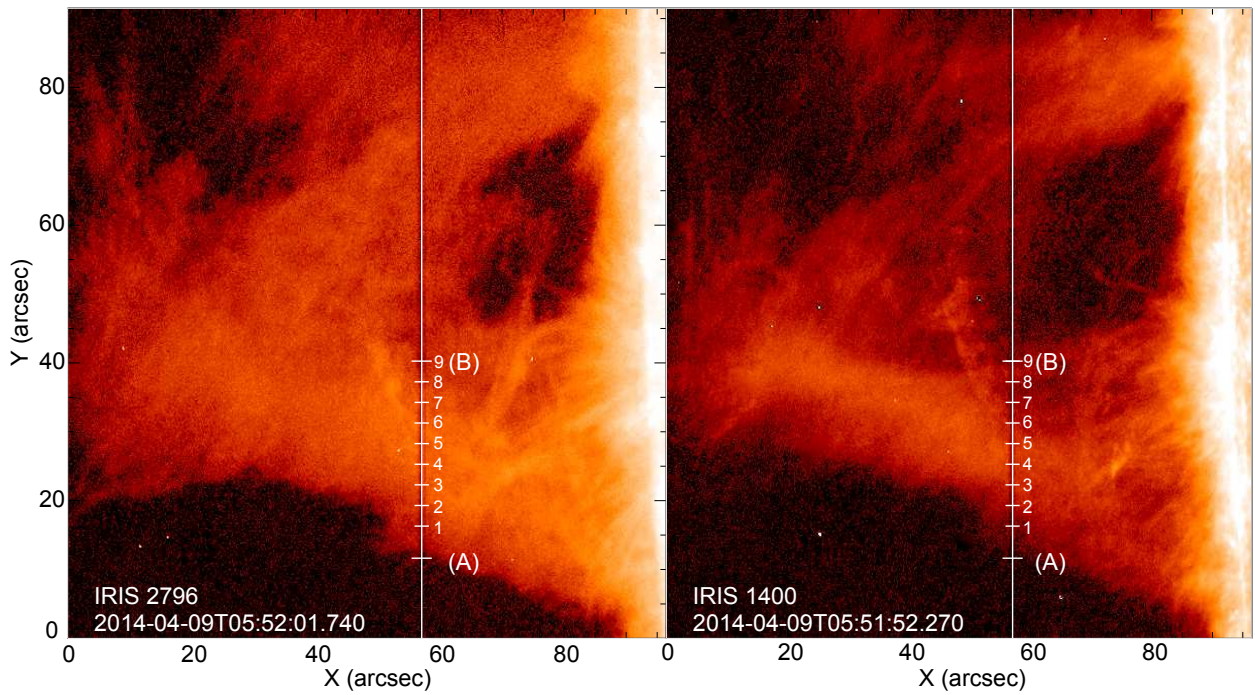}}
\caption{IRIS slit-jaw images of the 2796\AA{} (left) and 1400\AA{} (right) passbands for T1. The white vertical line in each image marks the location of the IRIS slit. The range between A and B is the range of the y-axis of Fig. 2. The numbers 1-9 indicate nine different positions where the line profiles are presented in Figures~\ref{fig.3} \& \ref{fig.4}. An animation is available online. }
\label{fig.1}
\end{figure*}

Figure~\ref{fig.1} and the associated online animation show the first solar tornado observed in the 2796\AA{} and 1400\AA{} passbands. The IRIS slit (white vertical line) crossed one leg of a prominence. The width of the leg is about 30$^{\prime\prime}$ ($\sim$22 Mm). The IRIS 2796\AA{} and 1400\AA{} passbands mainly sample the Mg~{\sc{ii}}~k~2796\AA{}~line formed at a temperature of about 0.01 MK and two Si~{\sc{iv}}~lines formed at a temperature of $\sim$0.1 MK, respectively. From the image sequences of both passbands, these seems to be apparent rotating motion of the prominence materials around the rotation axis. However, the rotation direction can not be easily inferred by just looking at the intensity images. Spectroscopic observations are required to determine whether  real rotation exists and what is the rotation direction in this prominence. 

To reveal the line-of-sight (LOS) velocities of this prominence leg or solar tornado, we have performed a single Gaussian fit (SGF) for the Si~{\sc{iv}}~1393\AA{}~line profiles observed in the tornado. The line profile of this optically thin line is generally Gaussian in the tornado. The  Mg~{\sc{ii}}~k~2796\AA{}~line is formed under the optically thick regime, and usually shows a reversal at the line core. But in prominences, the central reversal of Mg~{\sc{ii}}~k~is absent or very shallow \citep{Schmieder2014}, which is also the case in our observations. For this type of line profiles, the line centroid derived from the fitting should still reflect the velocity of the bulk motion. In addition, we are mainly interested in the change of Doppler shift across the tornado axis. Thus, a single Gaussian fit has also been applied to the profiles of this line observed in the tornado. The similar trend of Doppler shift found in both the Mg~{\sc{ii}}~and Si~{\sc{iv}}~lines also confirms the validity of this method. 

\begin{figure*}
\centering {\includegraphics[width=0.8\textwidth]{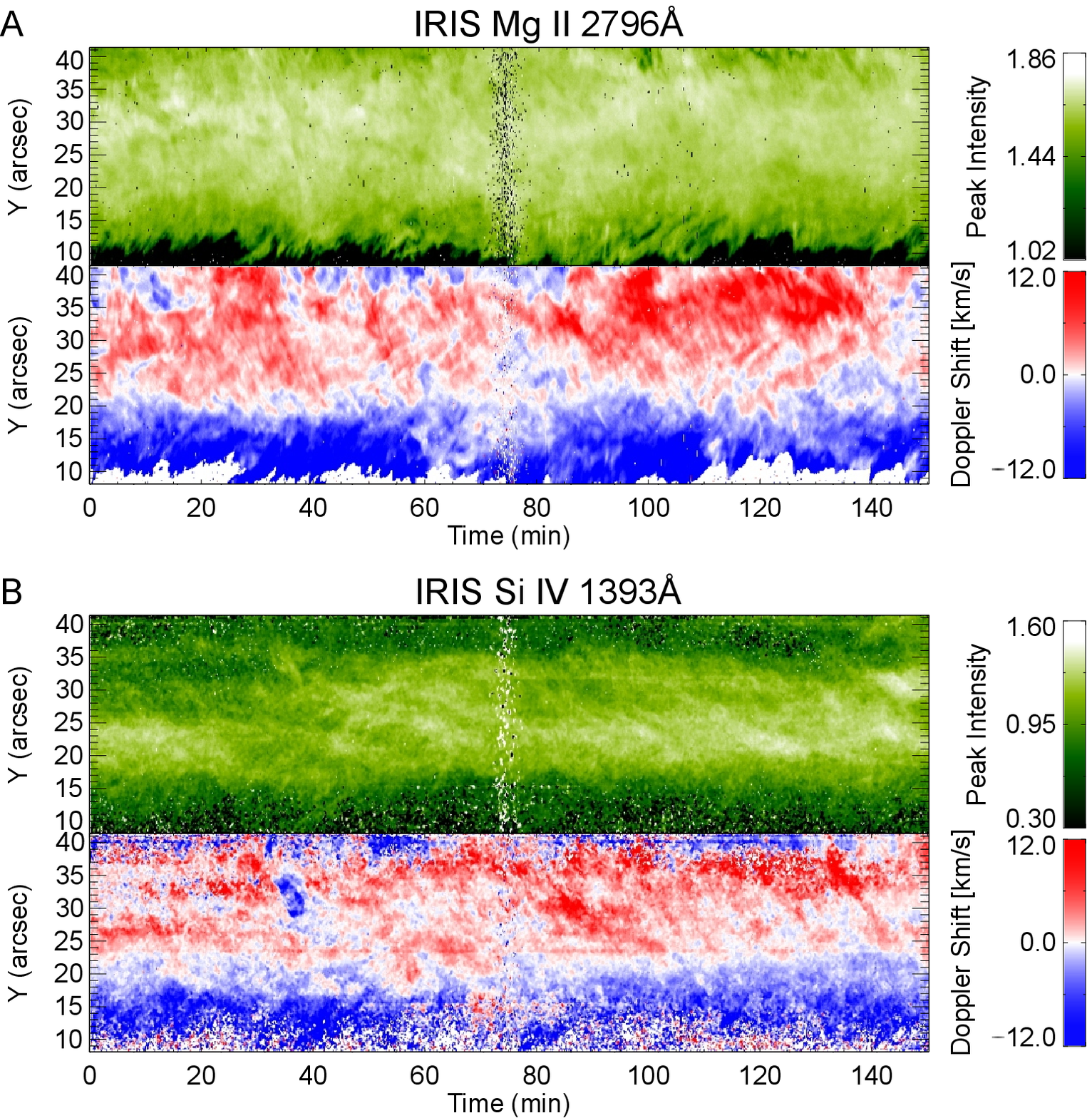}}
\caption{(A) Temporal evolution of the Doppler velocity and peak intensity of the Mg~{\sc{ii}}~k~2796\AA{}~line for T1. (B) Similar to (A) but for the Si~{\sc{iv}}~1393\AA{}~line. The peak intensities are shown in logarithmic scale.The range of the y-axis corresponds to the range between A and B as marked in Figure~\ref{fig.1}.}
\label{fig.2}
\end{figure*}

Through the SGF we have obtained the peak intensity and Doppler shift of each line at every pixel along the slit during the sit-and-stare observation. Figure~\ref{fig.2} shows the temporal evolution of the Doppler velocity and intensity along the slit for both the Mg~{\sc{ii}}~and Si~{\sc{iv}}~lines. The range of the y-axis is from A to B, across the tornado, as shown in Figure~\ref{fig.1}. A persistent and split red and blue shifts at two sides of the tornado can be seen from Figure~\ref{fig.2}. The absolute value of the maximum Doppler velocity is about 10~km~ s$^{-1}$. During the whole $\sim$2.5-hour observation period, the sign of the Doppler velocity is not changed on each side of the tornado.

\begin{figure*}
\centering {\includegraphics[width=\textwidth]{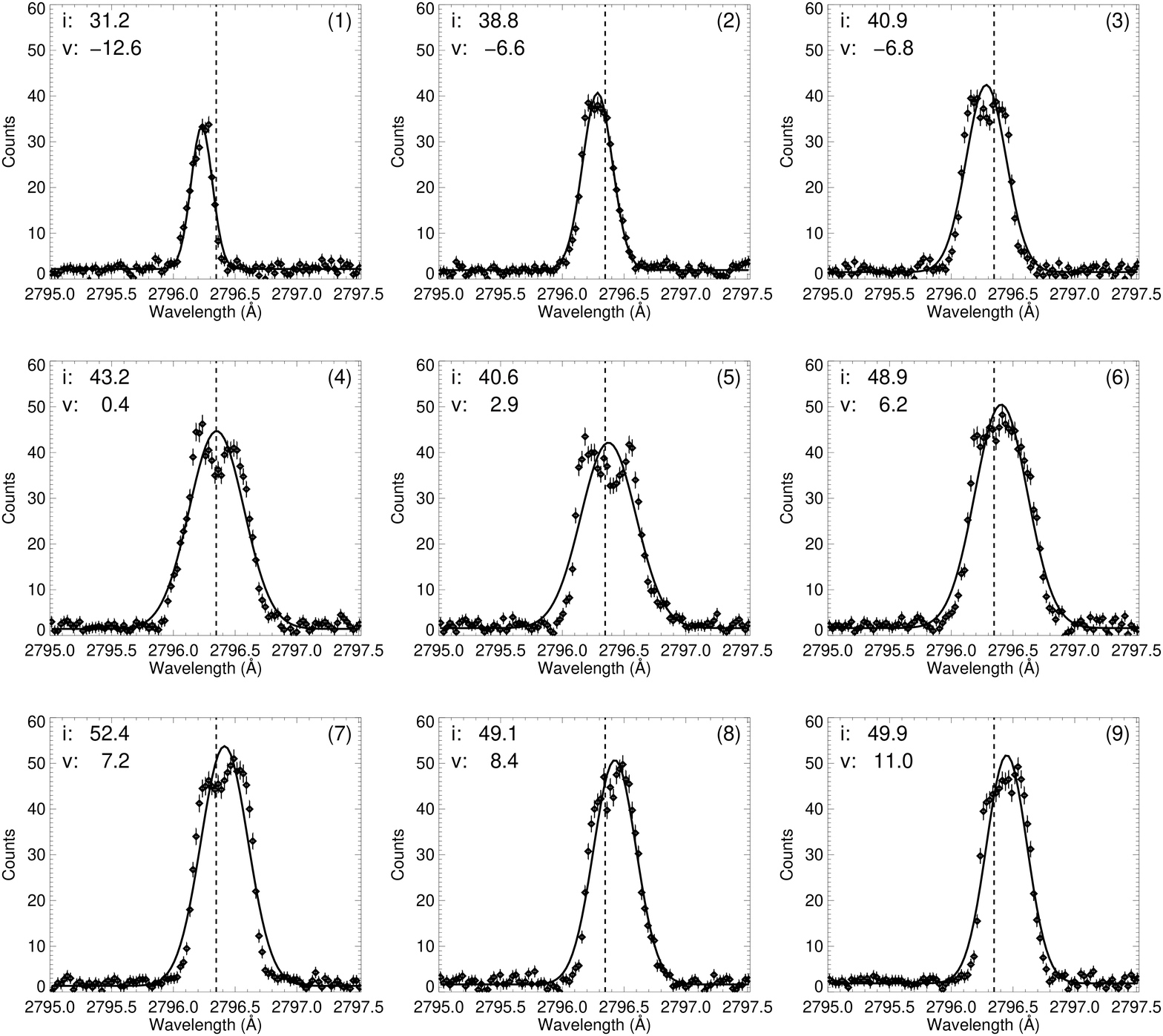}}
\caption{Mg~{\sc{ii}}~k~2796\AA{}~line profiles observed at nine positions (across the tornado, marked by the numbers 1-9 in Figure~\ref{fig.1}) along the IRIS slit. In each panel the diamonds and error bars represent the observed counts and measurement errors, respectively. We have added 2 to the observed counts for the purpose of illustration. The smooth line represents the Gaussian fit. The peak intensity (counts, "i") and Doppler velocity (km~ s$^{-1}$, "v") are also shown in each panel. The rest wavelength is indicated by the vertical dashed line. }
\label{fig.3}
\end{figure*}

\begin{figure*}
\centering {\includegraphics[width=\textwidth]{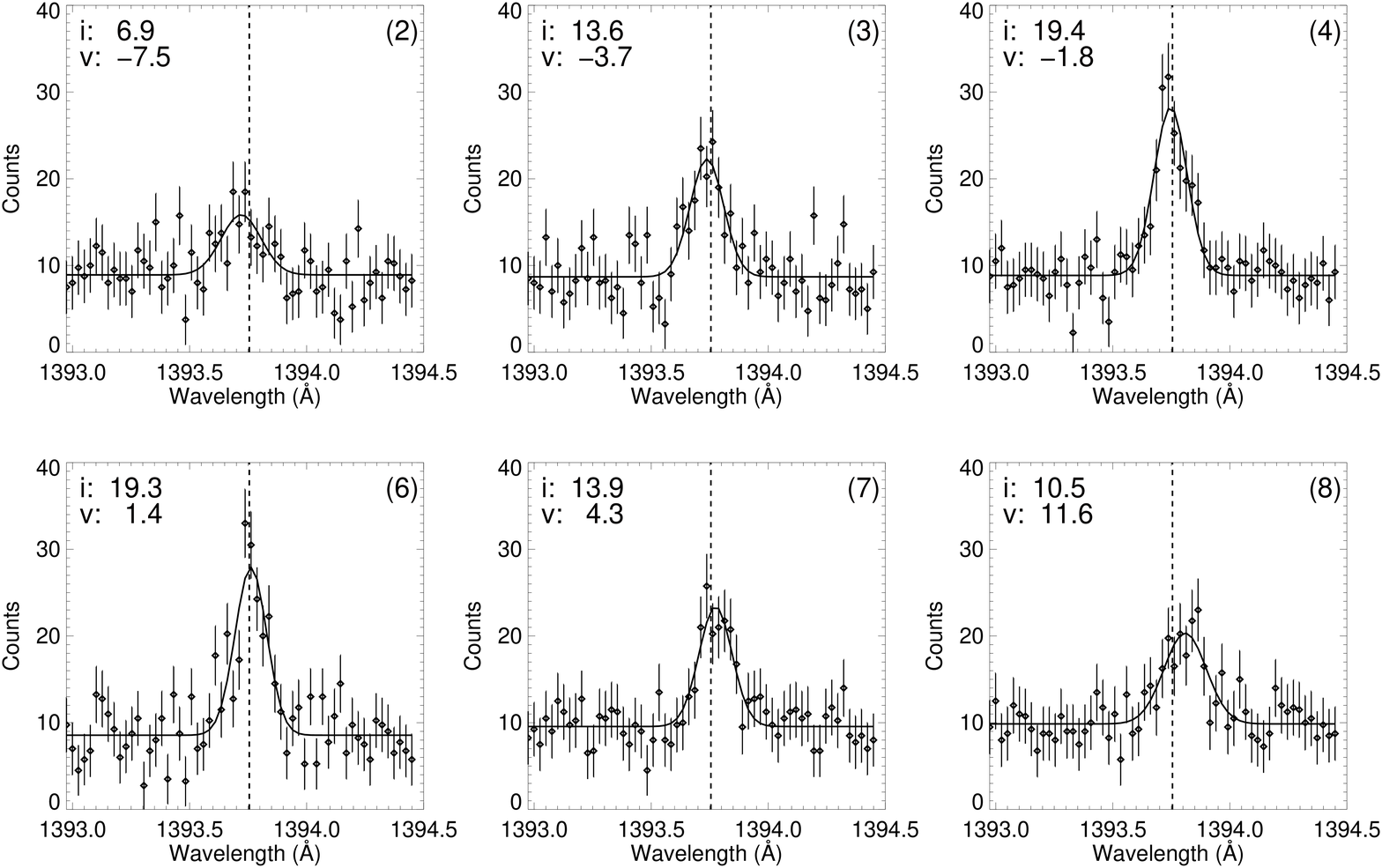}}
\caption{Similar to Fig. 3 but for the Si~{\sc{iv}}~1393\AA{}~line profiles at six positions marked in Figure~\ref{fig.1}. We have added 9 to the observed counts for the purpose of illustration. }
\label{fig.4}
\end{figure*}

Examples of the Mg~{\sc{ii}}~and Si~{\sc{iv}}~line profiles are presented in Figures~\ref{fig.3} \&~\ref{fig.4}. The measurement error of IRIS includes mainly two components: the photon counting error and CCD readout noise. The errors shown in the plots of line profiles are derived by adding the two in quadrature. The photon counting error is calculated as the square root of the photons. For the Si~{\sc{iv}}~line each DN corresponds to $\sim$4 photons and for the Mg~{\sc{ii}}~line one DN represents $\sim$18 photons. The readout noise is $\sim$3.1 DN for Si~{\sc{iv}} and 1.2 DN for Mg~{\sc{ii}}, which is related to the dark current uncertainty \citep{DePontieu2014}.

The nine Mg~{\sc{ii}}~profiles marked as 1-9  in Figures~\ref{fig.3} are observed at nine different positions along the slit (across the tornado, from one side to the other), as marked in Figure~\ref{fig.1}. The Doppler velocity changes from about --12~km~s$^{-1}$ (blue shift) on one side to nearly zero in the center of the tornado, and then to about 11~km~s$^{-1}$ (red shift) on the other side. 

Due to the weak Si~{\sc{iv}}~signal on the far sides of the tornado, we only plot line profiles at six positions across the tornado (Figure~\ref{fig.4}). The single Gaussian fit works well with the Si~{\sc{iv}}~1393\AA{}~line. Again the Doppler velocity changes from negative (blue shift) on one side of the tornado to positive (red shift) on the other side.

\begin{figure}
\centering {\includegraphics[width=0.5\textwidth]{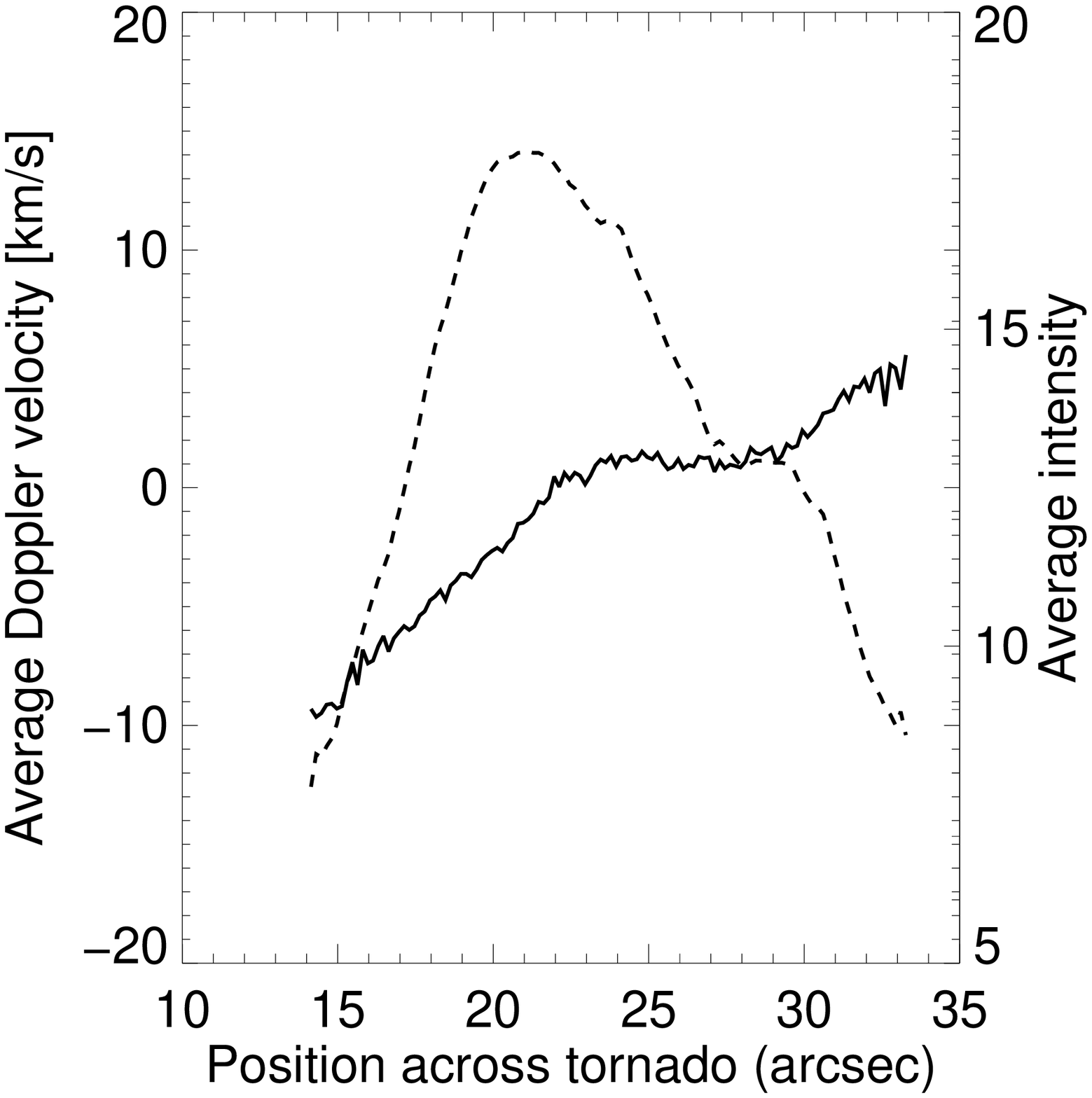}}
\caption{The change of average Doppler velocity (solid line) and peak intensity (dashed line) along the slit (across the tornado) for T1. The Doppler velocity and intensity are derived by applying a SGF to the Si~{\sc{iv}}~1393\AA{}~line profiles.}
\label{fig.5}
\end{figure}

Figure~\ref{fig.5} presents the change of the intensity and Doppler velocity of Si~{\sc{iv}}~across the tornado. The Doppler velocity (solid line) and peak intensity (dashed line) are averaged over the whole observation period. The average Doppler velocity gradually changes from negative (blue shift) on one side to positive (red shift) on the other side of the tornado. Meanwhile, the highest intensity appears near the center of the tornado, where the LOS velocity is $\sim$0~km~s$^{-1}$. Since the Si~{\sc{iv}}~line is optically thin, the intensity is proportional to the integrated emission along the LOS. Assuming a rotating cylinder or a helical flow/rotating plasma surrounding a central static structure with the uniform density, the integrated intensity should reach the peak value in the center as the integration length is the largest there; also, the velocity of the toroidal motion should have no component along the LOS at the center. The result presented in Figure~\ref{fig.5} appears to be consistent with this scenario. The gradual change of the LOS velocity appears to rule out the possibility of counter-streaming flows, as abrupt change of the velocity direction is expected at the boundary between the bidirectional flows. However, interactions between two oppositely directed flows may also lead to such a gradual change. Thus, additional observation signatures might be required to distinguish between the rotation and counter-streaming scenarios. 

\begin{figure}
\centering {\includegraphics[width=0.45\textwidth]{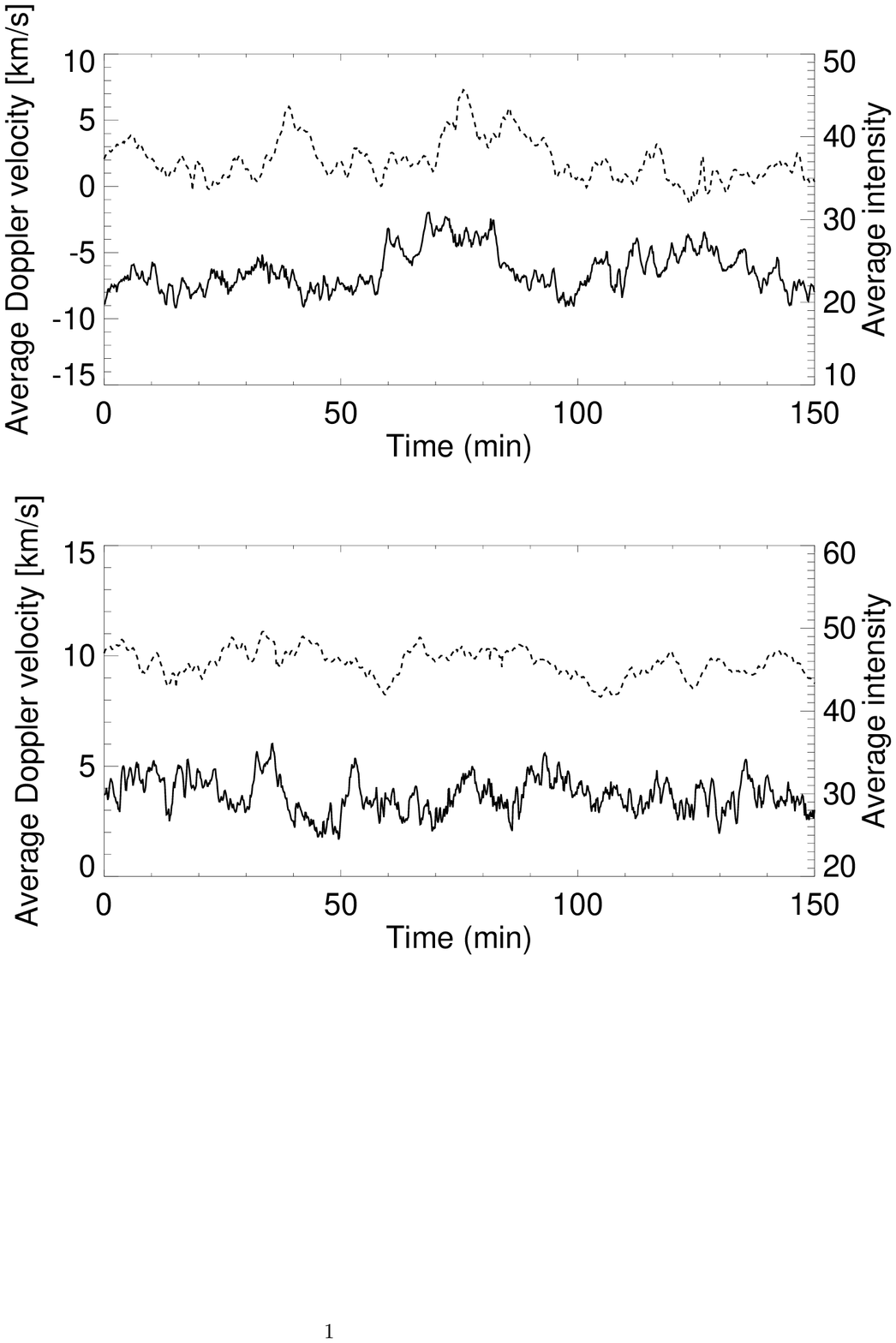}}
\caption{Temporal evolution of the average Doppler velocity (solid line) and intensity (dashed line) in the blueshifted (top panel) and redshifted (bottom panel) regions for T1. The Doppler velocity and peak intensity are derived by applying a SGF to the Si~{\sc{iv}}~1393\AA{}~line profiles.}
\label{fig.6}
\end{figure}

We also plot the temporal evolution of the average Doppler velocity and peak intensity of Si~{\sc{iv}}~1393\AA{}~on each side of the tornado. In Figure~\ref{fig.6}, the top panel shows the temporal evolution of the average Doppler shift (solid line) and intensity (dashed line) in the whole blueshifted region. The bottom panel presents the same for the redshifted region. During the 2.5-hour period of the observation, the average Doppler velocity in the blueshifted region remains negative, and that in the redshifted region remains positive. There is no sign change for the Doppler shift in either the blueshifted or redshifted region, which might not be easily explained by the oscillation scenario.

\subsection{Second tornado (T2)}
\begin{figure}
\centering {\includegraphics[width=0.45\textwidth]{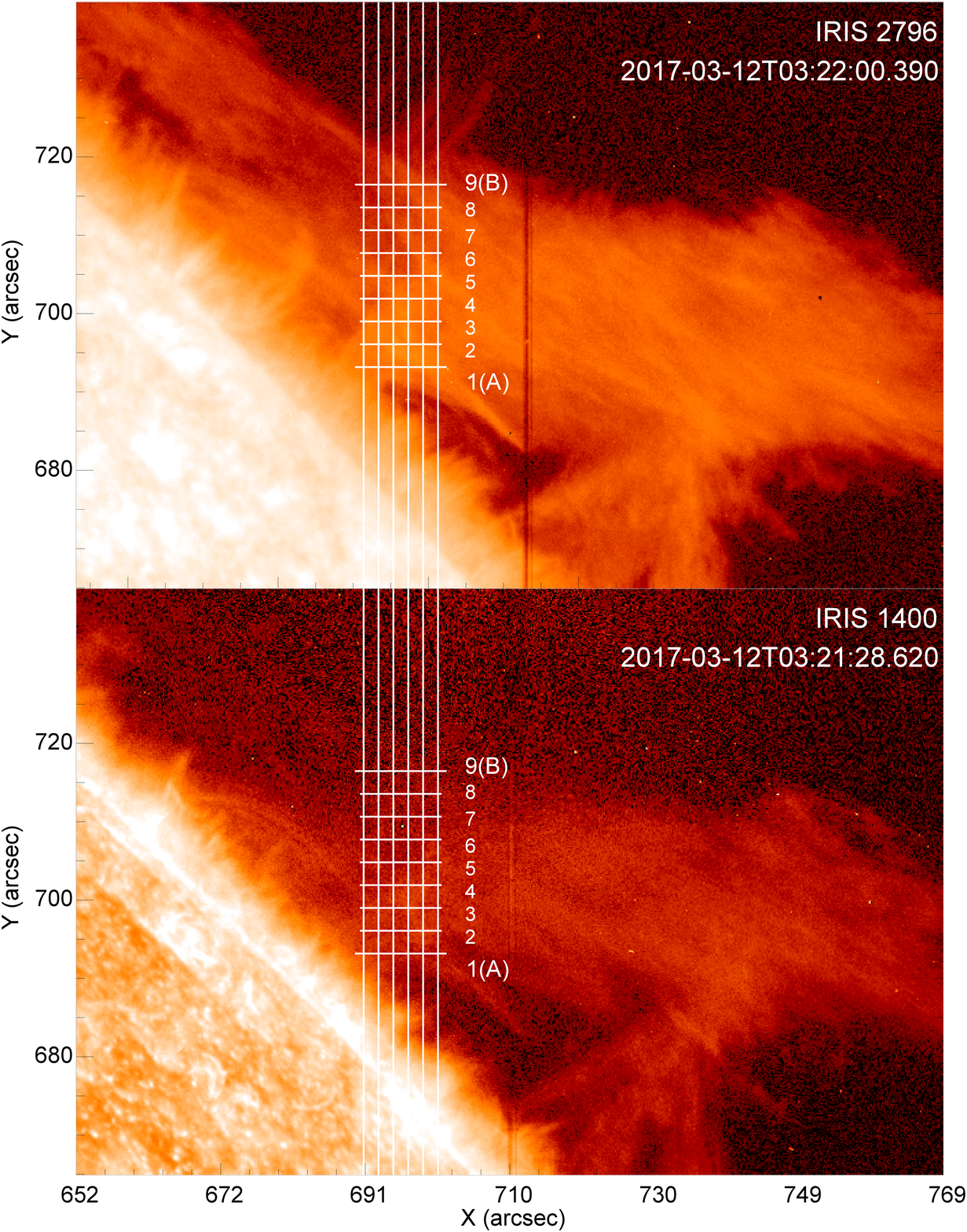}}
\caption{IRIS slit-jaw images in the 2796\AA{} (top) and 1400\AA{} (bottom) passbands for T2. The vertical white lines mark six slit locations during each raster scan. The range between A and B is the range of the y-axis of each panel in Figure~\ref{fig.8} \& \ref{fig.10}. The numbers 1-9 indicate nine different positions where the line profiles are presented in Figure~\ref{fig.9} \& \ref{fig.11}. An animation is available online.}
\label{fig.7}
\end{figure}

The way we process and analyze the data of the second solar tornado is basically the same as what we do for the first one. Figure~\ref{fig.7} and the associated online animation show the second solar tornado observed in the 1400\AA{} and 2796\AA{} passbands. The IRIS slit crossed one leg of a prominence, which appears to experience rotating motion around the rotation axis. The six white vertical lines that cross the solar tornado mark six selected slit locations from the 16-step repeated raster scans.

\begin{figure*}
\centering {\includegraphics[width=0.8\textwidth]{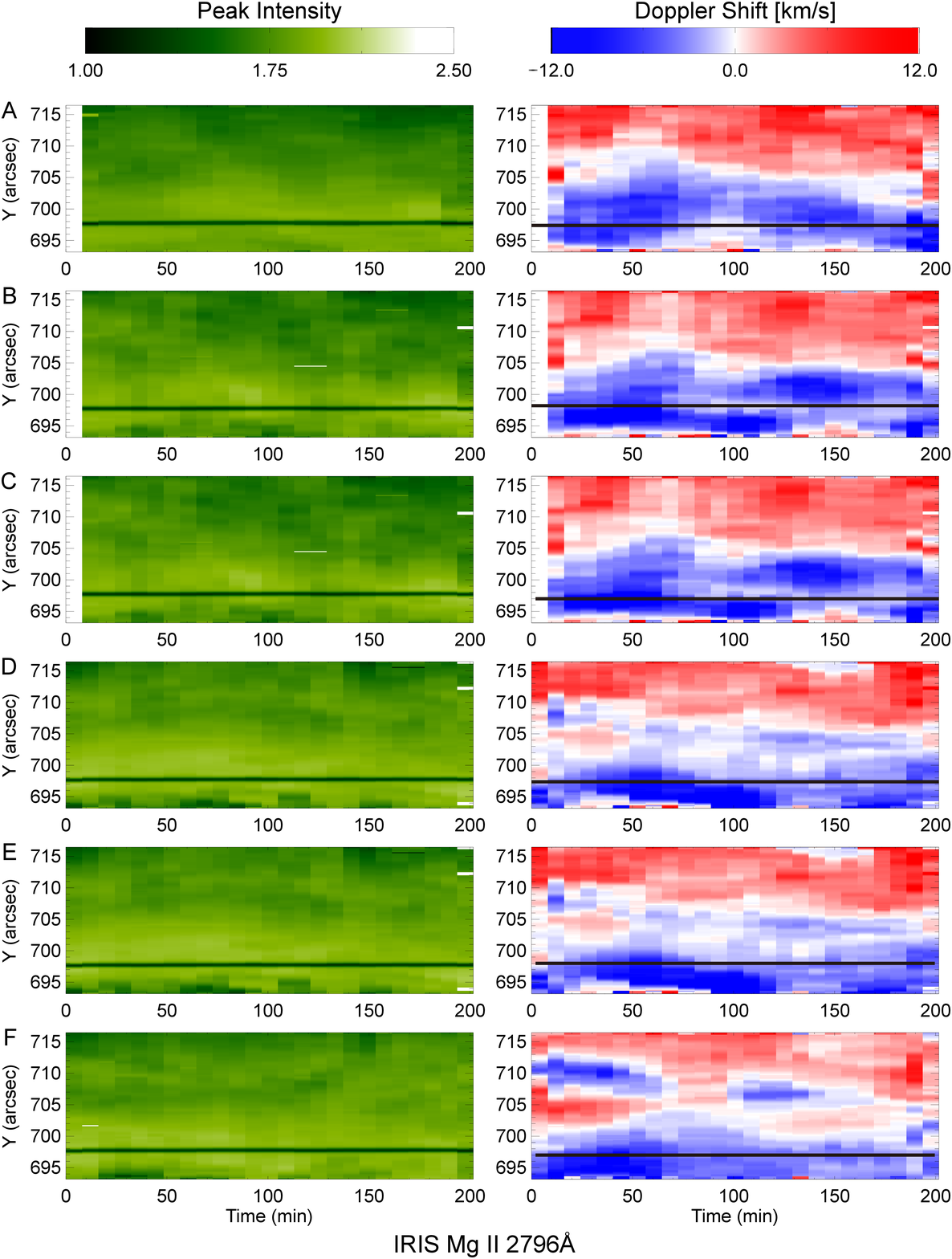}}
\caption{Temporal evolution of the Doppler velocity and peak intensity of the Mg~{\sc{ii}}~k~2796\AA{}~line for T2. The six rows show the results for the six different slit locations marked in Figure~\ref{fig.7}. The peak intensities are shown in logarithmic scale.The range of the y-axis in each panel corresponds to the range between A and B as marked in Figure~\ref{fig.7}. The dark horizontal lines in the images correspond to the fiducial mark on the IRIS slit. }
\label{fig.8}
\end{figure*}

Again we perform a single Gaussian fit to the Si~{\sc{iv}}~1393\AA{}~and Mg~{\sc{ii}}~k~2796\AA{}~line profiles, and obtain the Doppler velocities and peak intensities of the two emission lines at different spatial locations and different times. Figure~\ref{fig.8}(A)-(F) presents six sets of intensity and Doppler shift images of the Mg~{\sc{ii}}~k~line, corresponding to the six slit locations. Split red and blue shifts across the tornado axis can be found in all the six Dopplergrams. Such type of Doppler pattern is coherent and stable during the whole 3.5-hour period of the observation, possibly suggesting a coherent rotating motion of the cool tornado plasma.

\begin{figure*}
\centering {\includegraphics[width=\textwidth]{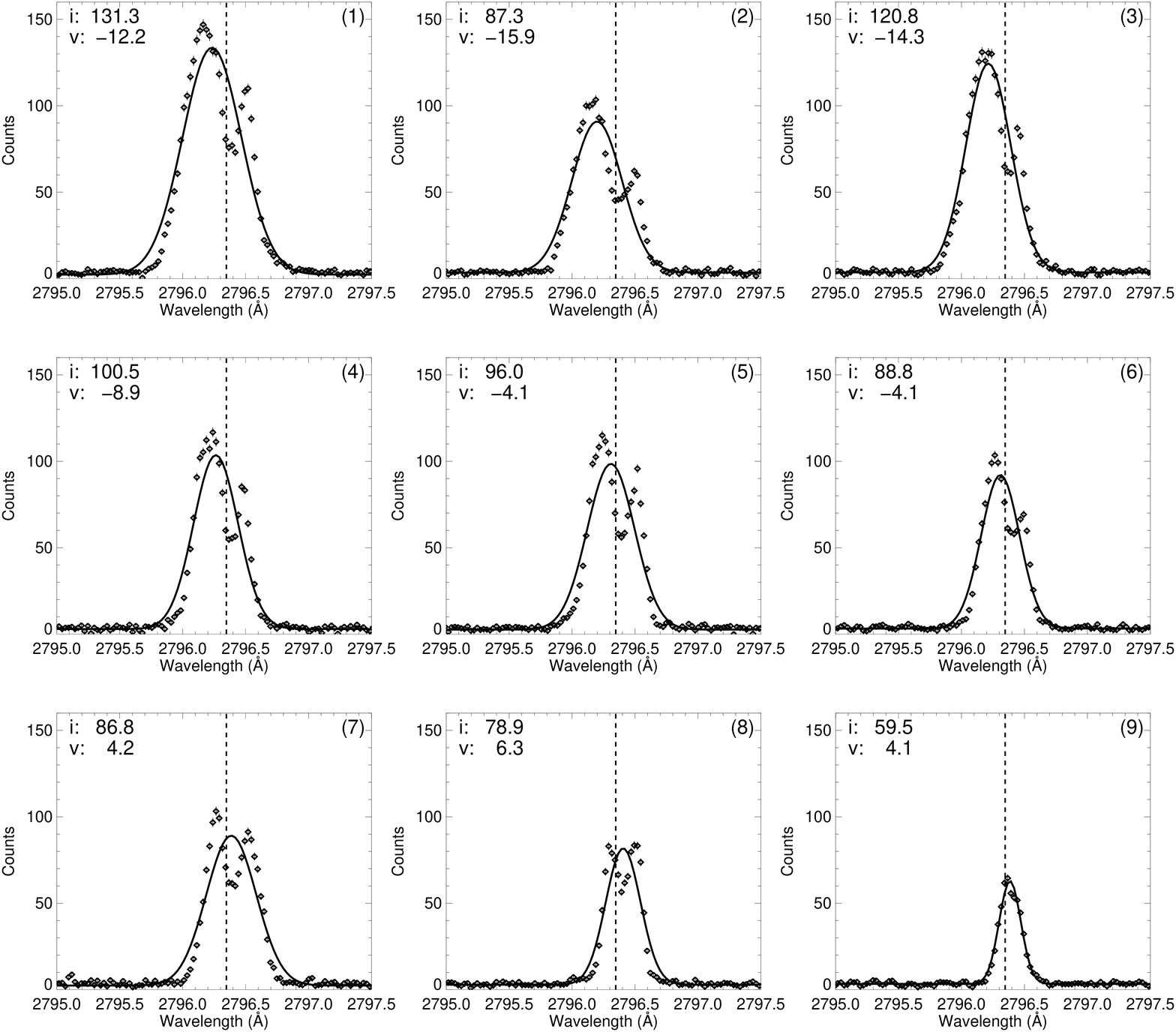}}
\caption{Mg~{\sc{ii}}~k~2796\AA{}~line profiles at nine positions (marked by the numbers 1-9 in Figure~\ref{fig.7}) along the IRIS slit (across the tornado) in the observation of T2. In each panel the diamonds and error bars represent the observed counts and measurement errors, respectively. We have added 2 to the observed counts for the purpose of illustration. The smooth line represents the Gaussian fit. The peak intensity (counts, "i") and Doppler velocity (km~ s$^{-1}$, "v") are also shown in each panel. The rest wavelength is indicated by the vertical dashed line.}
\label{fig.9}
\end{figure*}

We present several profiles of the Mg~{\sc{ii}}~line and their fitting results in Figure~\ref{fig.9}. These nine profiles were observed at nine different locations across the tornado, from one side to the other. Consistent with the Doppler signature depicted in Figure~\ref{fig.8}, the Doppler velocity changes from blueshift to redshift across the solar tornado.

\begin{figure*}
\centering {\includegraphics[width=0.8\textwidth]{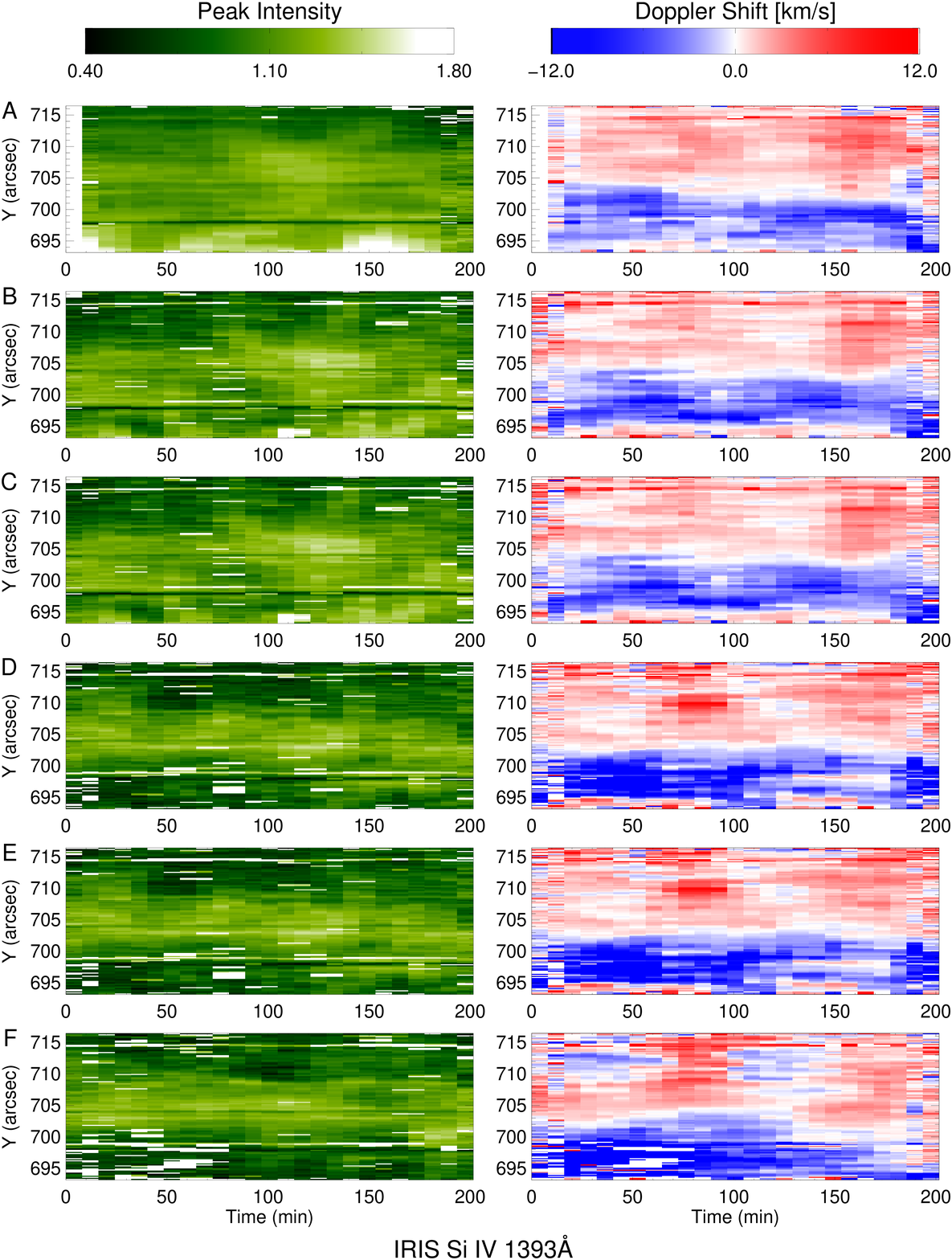}}
\caption{Similar to Figure~\ref{fig.8} but for the Si~{\sc{iv}}~1393\AA{}~line.}
\label{fig.10}
\end{figure*}

\begin{figure*}
\centering {\includegraphics[width=\textwidth]{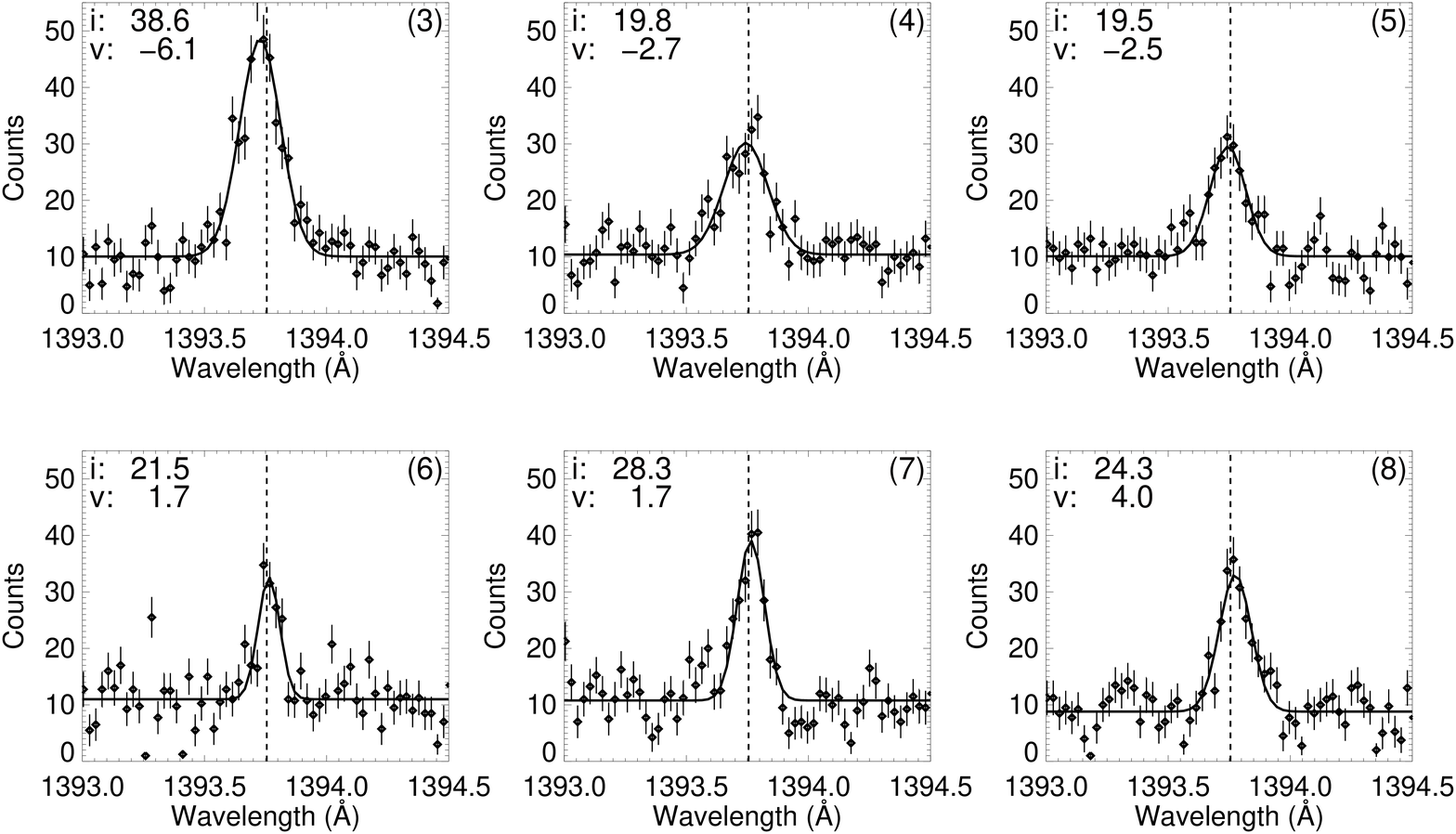}}
\caption{Similar to Figure~\ref{fig.9}  but for the Si~{\sc{iv}}~1393\AA{}~line profiles at six positions marked in Figure~\ref{fig.7}. We have added 9 to the observed counts for the purpose of illustration.}
\label{fig.11}
\end{figure*}

The intensity and Doppler shift images for the Si~{\sc{iv}}~1393\AA{} are presented in Figure~\ref{fig.10}. Similar to Figure~\ref{fig.8}, the Si~{\sc{iv}}~Dopplergrams also reveal adjacent red and blue shifts across the axis of the tornado, without any change of sign throughout the observation. In Figure~\ref{fig.11}, six line profiles obtained from six different positions across the tornado reveal a similar trend of the Doppler shift: from blue shift on one side of the tornado to red shift on the other side.

\begin{figure*}
\centering {\includegraphics[width=\textwidth]{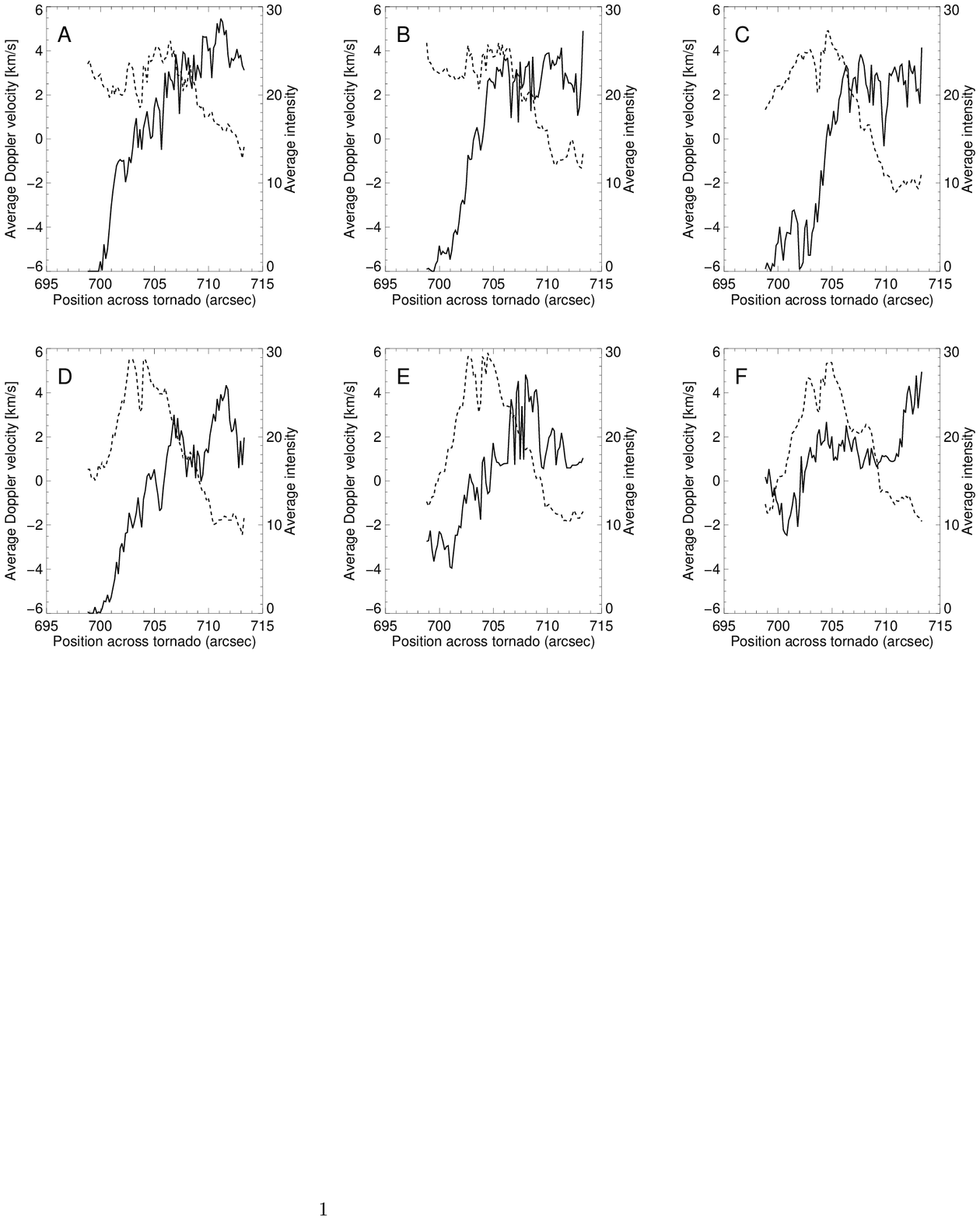}}
\caption{The change of average Doppler velocity (solid line) and intensity (dashed line) along the slit (across the tornado) for T2. The Doppler velocity and intensity are derived by applying a SGF to the Si~{\sc{iv}}~1393\AA{}~line profiles. The six panels represent results derived at the six slit locations marked in Fig.~\ref{fig.7}.}
\label{fig.12}
\end{figure*}

We also present the average Doppler velocity (solid line) and intensity (dashed line) of the Si~{\sc{iv}}~line along the slit in Figure~\ref{fig.12}. At each spatial pixel, the Doppler velocity and peak intensity are averaged during the whole observing period. The six panels correspond to the six slit locations. At all the six slit locations, the average Doppler velocity exhibits the trend that it changes from blue shift on one side of the tornado to red shift on the other side. The maximum value of the average intensity is found at the location where the velocity is $\sim$0~km~s$^{-1}$ (the center of the solar tornado). As stated in Section 3.1, this result is consistent with the scenario of a rotating tornado.

\begin{figure*}
\centering {\includegraphics[width=\textwidth]{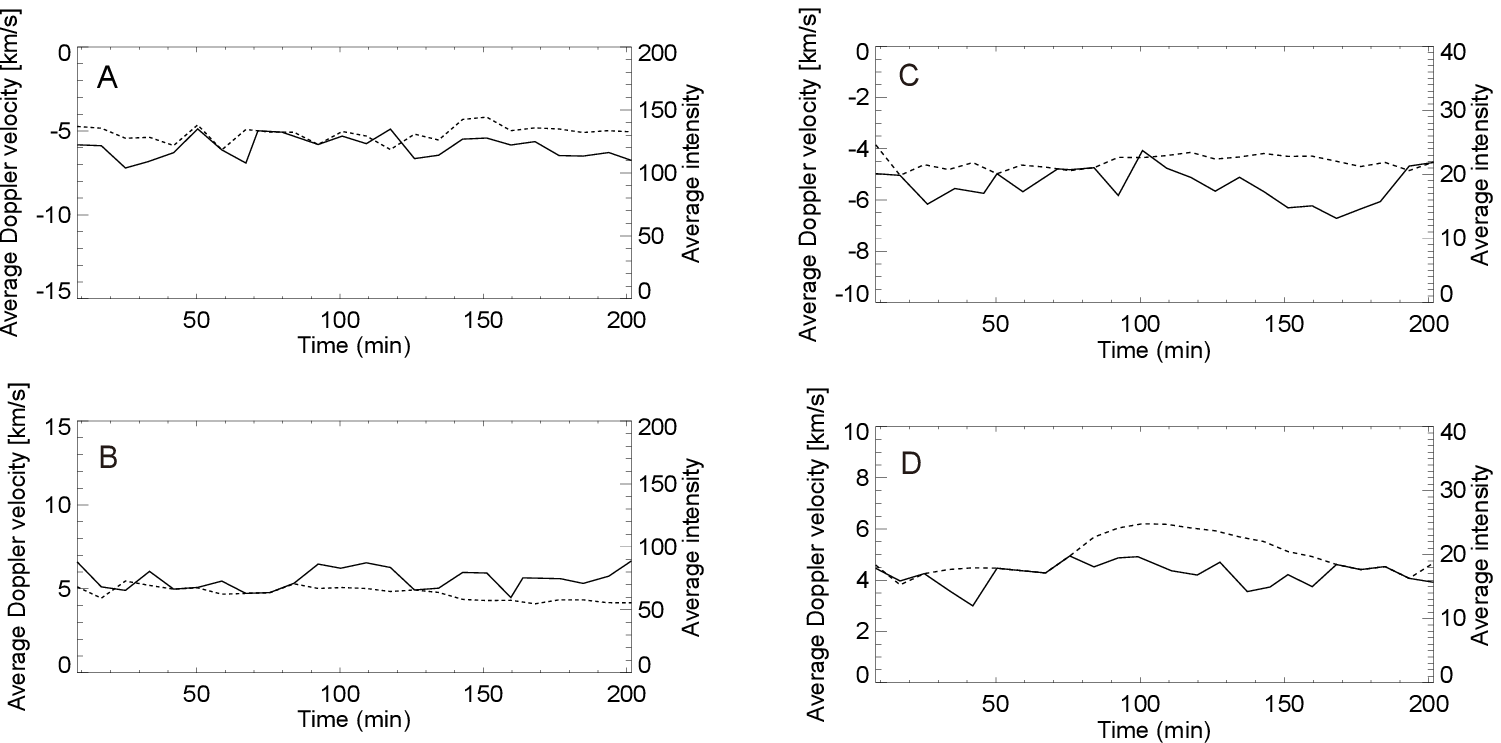}}
\caption{Temporal evolution of the average Doppler velocity (solid line) and intensity (dashed line) in the blueshifted (top) and redshifted (bottom) regions at the first selected slit location (see Figure~\ref{fig.7}) for T2. The Doppler velocity and peak intensity are derived from SGF to the Mg~{\sc{ii}}~k~2796\AA{}~(A-B) and Si~{\sc{iv}}~1393\AA{}~(C-D) line profiles.}
\label{fig.13}
\end{figure*}

Figure~\ref{fig.13} shows the temporal evolution of the average Doppler velocity (solid line) and intensity (dashed line) in the blueshifted and reshifted regions for the Mg~{\sc{ii}}~and Si~{\sc{iv}}~lines. Again, no sign reversal can be identified from the Doppler shift on either side of the tornado.

\begin{figure*}
\centering {\includegraphics[width=\textwidth]{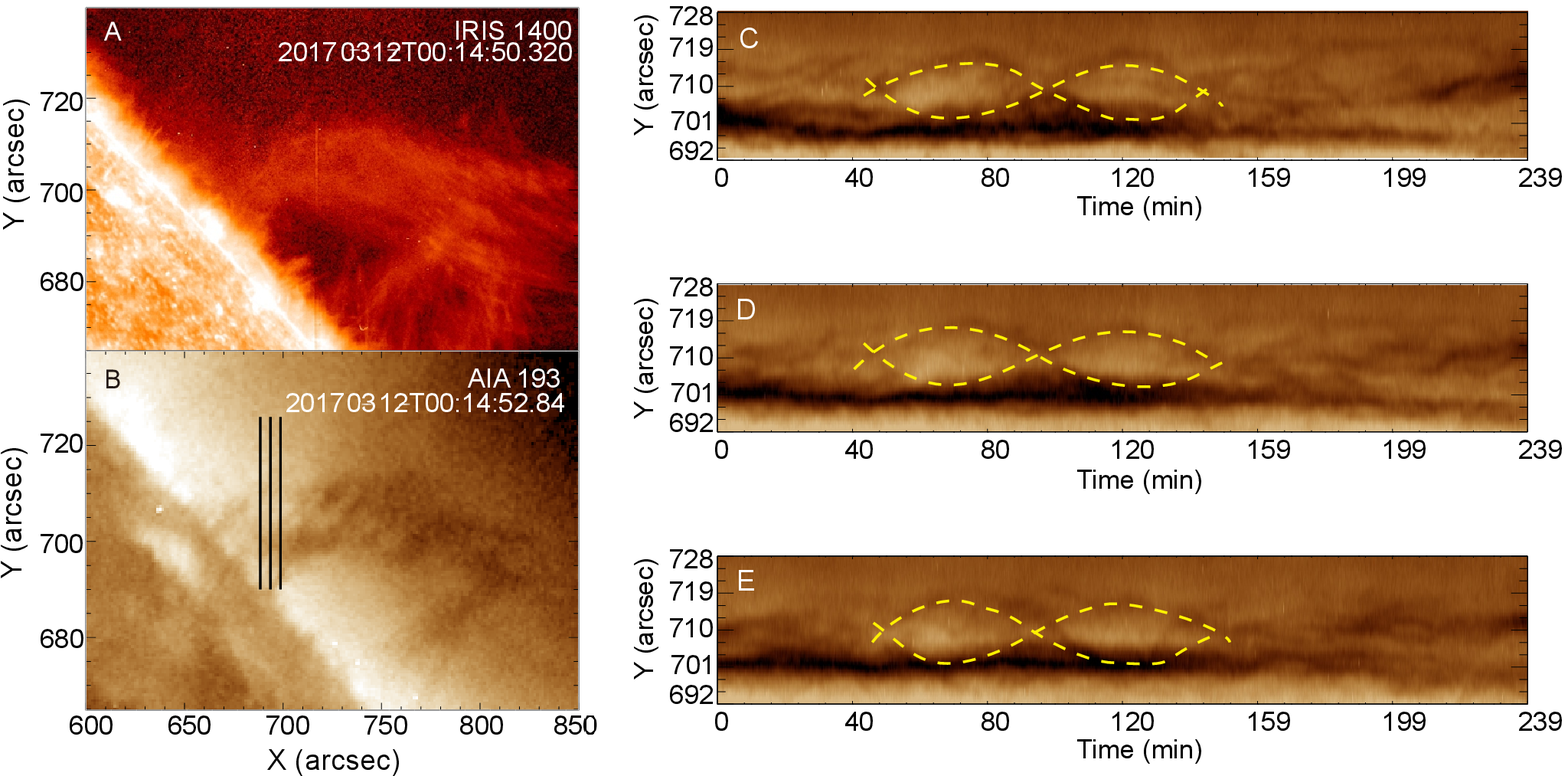}}
\caption{ (A)-(B): An IRIS 1400\AA{} image (top) coaligned with an AIA 193\AA{} image (bottom). The three dark vertical lines indicate the first three slit locations as shown in Figure~\ref{fig.7}. (C)-(E): Space-time diagrams for these three lines. The yellow dashed lines mark crossed dark structures, indicating periodic motion. An animation showing the image sequence of AIA 193\AA{} is available online. }
\label{fig.14}
\end{figure*}

We have also analyzed the images taken with SDO/AIA for the second solar tornado. Figure~\ref{fig.14}(C)-(E) shows three space-time diagrams of the AIA 193\AA{} intensity. The three diagrams are obtained at the first three slit locations of IRIS as shown on the left panel of Figure~\ref{fig.14}. We obtain a result similar to that of \cite{Su2014}. Although the motion of the darkest feature is hard to track, we can still find crossing dark structures with less absorption. The crossed dark structures on each space-time diagram, as marked by the yellow dashed lines, are a sign of periodic motion. Such crossed structures can be caused either by the rotation of cold materials or periodic oscillation, but it may not be easily explained by counter-streaming. Because the oscillation scenario is likely not consistent with our spectroscopic observations, this result can only be explained by the rotation of cool materials in the tornado.

\section{Summary}

Previous analyses of solar tornadoes led to different interpretations about the apparent motion of the tornadoes: rotation, oscillation, and counter-streaming. With spectroscopic observations of the cool tornado materials by IRIS, we demonstrate that the apparent rotating motion of the two tornadoes studied here is most likley caused by the rotating plasma in the tornadoes.

The Doppler shifts of the Mg~{\sc{ii}}~k~2796\AA{}~and the Si~{\sc{iv}}~1393\AA{}~lines reveal crucial information about the nature of the motion in solar tornadoes. Split red and blue shifts across the tornado axis are found in both observations. In the second observation, a similar Doppler pattern is found at different locations along the tornado axis. A similar Doppler pattern for hot coronal lines has been previously found by \cite{Su2014} and \cite{Levens2015}, and was interpreted as rotating motion of the million-degree plasma surrounding the tornado. The Doppler shifts we obtained reflect the mass motion of the cool tornado materials, since the Mg~{\sc{ii}}~and Si~{\sc{iv}}~lines are formed at a temperature of $\sim$0.01 MK and $\sim$0.1 MK, respectively.

Both observations lasted for more than 2.5 hours, a duration that is long enough to identify possible periodic oscillation (alternation of red and blue shifts) with a $\sim$1-hour period as identified by \cite{Schmieder2017}. However, during the whole observations, these adjacent red and blue shifts appear to be persistent, and they do not show any sign change of the Doppler shift. Thus, the apparent rotating motion in our observations is likely not dominated by oscillations. However, we may not rule out the possibility of oscillations in tornado structures. For instance, some type of oscillations with relatively small amplitudes could be superimposed on the overwhelming rotating motion.

In both observations, the average Doppler velocity changes gradually from the blue shift on one side of the tornado to nearly zero in the center, and then to red shift on the other side. The gradual change of the Doppler shift across the tornado axis might not be easily explained by counter-streaming flows, unless there are strong interactions between the oppositely directed flows. The maximum intensity of the Si~{\sc{iv}}~line is generally found around the center of the tornado, where the LOS velocity is nearly zero. Such behavior is consistent with the scenario of rotating plasma.

The AIA observation of the second tornado may rule out the possibility of counter-streaming flows. The crossed dark structures in the space-time diagrams of AIA 193\AA{} intensity are a sign of periodic motion: either rotation of cool plasma or oscillation of plasma in solar tornadoes. It appears difficult to explain this periodic motion by counter-streaming flows.

In summary, analysis results from our IRIS and AIA observations of solar tornadoes appear to provide evidence to the viewpoint of cool rotating materials in the two tornadoes studied here. There may be two different scenarios: the entire magnetic structure of a tornado is rotating \citep[magnetic tornado,][]{Su2012}, or cool materials are flowing along a static helical magnetic structure. If the entire magnetic structure of a tornado is rotating, we can estimate the time used for the coronal field lines to be dragged by one turn, which is $\sim$1.3 hours if we use a rotation speed of $\sim$10~km~s$^{-1}$ and a width of $\sim$20$^{\prime\prime}$. Considering the fact that the two tornadoes we study here are seen to exist for around a week close to the solar limb from AIA observations, the magnetic field lines of these tornadoes would be wound around by more than 100 turns. Such highly twisted magnetic structures are obviously unstable and would inevitably lead to eruptions before so many rotations. Thus, continuous rotation of the entire magnetic structures associated with the tornadoes appears to have difficulty in explaining the observed long lifetimes of the tornadoes. While the scenario of cool materials flowing along a relatively stable helical magnetic structure is not in contradiction with the existence of these structures for weeks. However, due to the lack of precise measurement of the magnetic fields in the tornadoes, we cannot exactly comment if this scenario is correct or not.

\begin{acknowledgements}
IRIS is a NASA small explorer mission developed and operated by LMSAL with mission operations executed at NASA Ames Research center and major contributions to downlink communications funded by ESA and the Norwegian Space Centre. SDO is a mission in NASA's Living With a Star (LWS) Program. This work is supported by the Recruitment Program of Global Experts of China, the Max-Planck Partner Group program,  NSFC under grants 41574166, 11790304 and 11790300, and the Joint Research Fund in Astronomy (U1631242) under cooperative agreement between the NSFC and CAS. We thank Dr. Brigitte Schmieder and Prof. Pengfei Chen for helpful discussions.
\end{acknowledgements}

{}

\begin{thebibliography}{}
\bibitem[Chen et al.(2017)]{Chen2017}
Chen, H., Zhang, J., Ma, S., et al. 2017, ApJL, 841, L13

\bibitem[Culhane et al.(2007)]{Culhane2007}
Culhane, J. L., Harra, L. K., James, A. M., et al. 2007, Sol. Phys., 243, 19

\bibitem[Curdt \& Tian(2011)]{Curdt&Tian2011}
Curdt, W., \& Tian, H. 2011, A\&A, 532, L9

\bibitem[De Pontieu et al.(2014)]{DePontieu2014}
De Pontieu, B., Title, A. M., Lemen, J. R., et al. 2014, Sol. Phys., 289, 2733

\bibitem[Gonz\'alez et al.(2016)]{Gonzalez2016}
Gonz\'alez, M. M., Ramos, A. A., Arregui, I., et al. 2016, ApJ, 825, 119

\bibitem[Lemen et al.(2011)]{Lemen2012}
Lemen, J. R., Title, A. M., Akin, D. J., et al. 2012, Sol. Phys., 275, 17

\bibitem[Levens et al.(2015)]{Levens2015}
Levens, P. J., Labrosse, N., Fletcher, L., Schmieder, B., 2015, A\&A, 582, A27

\bibitem[Levens et al.(2016a)]{Levens2016a}
Levens, P. J., Schmieder, B., Labrosse, N., L\'opez Ariste, A. 2016a, ApJ, 818, 31

\bibitem[Levens et al.(2016b)]{Levens2016b}
Levens, P. J., Schmieder, B., L\'opez Ariste, A., et al. 2016b, ApJ, 826, 164

\bibitem[Levens et al.(2017)]{Levens2017}
Levens, P. J., Labrosse, N., Schmieder, B., L\'opez Ariste, A., \& Fletcher, L. 2017, A\&A, 607, A16

\bibitem[Li et al.(2012)]{Li2012}
Li, X., Morgan, H., Leonard, D., \& Jeska, L. 2012, ApJL, 752, L22

\bibitem[Luna et al.(2015)]{Luna2015}
Luna, M., et al. 2015, ApJL, 808, L23

\bibitem[Martin(1998)]{Martin1998}
Martin, S. F. 1998, Sol. Phys., 182, 107

\bibitem[Panasenco et al.(2014)]{Panasenco2014}
Panasenco, O., Martin, S. F., \& Velli, M. 2014, Sol. Phys., 289, 603

\bibitem[Panesar et al.(2013)]{Panesar2013}
Panesar, N. K., Innes, D. E., Tiwari, S. K., \& Low, B. C. 2013, A\&A, 549, A105

\bibitem[Parenti(2014)]{Parenti2014}
Parenti, S. 2014, LRSP, 11, 1

\bibitem[Pesnell et al.(2012)]{Pesnell2012}
Pesnell, W. D., Thompson, B. J., \& Chamberlin, P. C. 2012, Sol. Phys., 275, 3

\bibitem[Pettit(1943)]{Pettit1943}
Pettit, E. 1943, ApJ, 98, 6

\bibitem[Pike \& Mason(1998)]{Pike&Mason1998}
Pike, C. D., \& Mason, H. E. 1998, Sol. Phys., 182, 333

\bibitem[Schmieder et al.(2014)]{Schmieder2014}
Schmieder, B., Tian, H., Kucera, T., et al. 2014, A\&A, 569, A85

\bibitem[Schmieder et al.(2017)]{Schmieder2017}
Schmieder, B., Mein, P., Mein, N., et al. 2017, A\&A, 597, A109

\bibitem[Shen et al.(2011)]{Shen2011}
Shen, Y., Liu, Y., Su, J., \& Ibrahim, A. 2011, ApJL, 735, L43

\bibitem[Shen et al.(2012)]{Shen2012}
Shen, Y., Liu, Y., Su, J. et al. 2012, ApJ, 745, 164

\bibitem[Shen et al.(2015)]{Shen2015}
Shen, Y., Liu, Y., Liu, Y. D., et al. 2015, ApJ, 814, L17

\bibitem[Su et al.(2012)]{Su2012}
Su, Y., Wang, T., Veronig, A., Temmer, M., \& Gan, W. 2012, ApJL, 756, L41

\bibitem[Su et al.(2014)]{Su2014}
Su, Y., G\"om\"ory, P., Veronig, A., et al. 2014, ApJL, 785, L2

\bibitem[Tian et al.(2014)]{Tian2014}
Tian, H., DeLuca, E., Reeves, K. K., et al. 2014, ApJ, 786, 137

\bibitem[Wedemeyer-B\"ohm \& Rouppe van der Voort(2009)]{Wedemeyer&Voort2009}
Wedemeyer-B?hm, S., \& van der Voort, L. R. 2009, A\& A, 507, L9

\bibitem[Wedemeyer-B\"ohm et al.(2012)]{Wedemeyer2012}
Wedemeyer-B\"ohm, S., Scullion, E., Steiner, O., et al. 2012, Nature, 486, 505

\bibitem[Wedemeyer-B\"ohm et al.(2013)]{Wedemeyer2013}
Wedemeyer-B\"ohm, S., Scullion, E., van der Voort, et al. 2013, ApJ, 774, 123

\bibitem[Yang et al.(2015)]{Yang2015}
Yang, B., Jiang, Y., Yang, J., et al. 2015, ApJ, 803, 86

\bibitem[Zhang \& Liu(2011)]{Zhang&Liu2011}
Zhang, J., \& Liu, Y. 2011, ApJL, 741, L7
\end{thebibliography}
\end{document}